\documentclass[superscriptaddress,altaffilletter,amssymb,showpacs,nofootinbib]{revtex4}
\usepackage[dvips]{graphicx}
\usepackage{amsmath}
\newcommand{\be}{\begin{equation}}
\newcommand{\ee}{\end{equation}}
\newcommand{\bea}{\begin{eqnarray}}
\newcommand{\eea}{\end{eqnarray}}

\begin{document}

\title{Exact models with non-minimal interaction between dark matter\\ and (either phantom or quintessence) dark energy}

\author{Tame Gonzalez}\email{tame@uclv.edu.cu}
\author{Israel Quiros}\email{israel@uclv.edu.cu}
\affiliation{Universidad Central de Las Villas, Santa Clara CP 54830, Cuba}

\date{\today}
\begin{abstract}
A method for deriving Friedmann-Robertson-Walker (FRW) solutions developed in Int. J. Mod. Phys. D{\bf 5}(1996)71-84, is
generalized to account for models with non-minimal coupling between the dark energy and the dark matter. New quintessence and phantom (flat) FRW solutions are found. Their physical significance is discussed. Additionally, the aforementioned method is modified so that, "coincidence free" solutions can be readily derived. Besides, we review some aspects of the phantom barrier crossing. In this regard we present a model which is free from the coincidence problem and, at the same time, does the crossing of the phantom barrier $\omega=-1$ at late time. Finally, we give additional comments on the non predictive properties of scalar field cosmological models with or without energy transfer. 
\end{abstract}


\maketitle


\section{Introduction}

Although many alternative models of the universe that take account of the present stage of accelerated expansion are being studied, due to their simplicity dark energy (DE) models are perhaps the preferred ones in the bibliography. Notwithstanding these are plagued by many problems, among which is the so called "cosmic coincidence" (CC) problem. Other undesirable features of these models are related with the possibility- already anticipated by the observational data- that the DE equation of state (EOS) parameter might be more negative than minus unity (the limiting value of the vacuum EOS parameter). This led cosmologists to face the so called "phantom" dark energy, i.e., a DE component with negative kinetic energy term. No matter whether or not there are real chances for the dark energy (whether vacuum, quintessence or phantom) to exist, the study of such models is very interesting and it could hint at more compelling possibilities.

Perhaps, the simplest realization of the dark energy is a minimally coupled (self-interacting) scalar field. Due to their mathematical simplicity, these models- also called as "quintessence" or "decaying vacuum" DE- have been intensively and detailed studied over decades. The next step towards a more complicated and, may be, more realistic model, is to consider the possibility of an additional non gravitational interaction between the DE and the background cosmic fluid (see reference \cite{tsujikawa1} for a general formalism to describe the coupled quintessence, phantom, non phantom K-essence and Tachyon, etc, and \cite{tsujikawa2} for a coherent review). Otherwise the scalar field DE is coupled non-minimally to the background fluid (usually cold dark matter) and, consequently, the background particles do not follow the geodesics of the metric, i.e, these are coupled to a scalar-tensor metric (see below for details).

Although experimental tests in the solar system impose severe restrictions on the possibility of non-minimal coupling between the DE and ordinary matter fluids (the background)\cite{will}, due to the unknown nature of the dark matter (DM) as the major part of the background, it is possible to have additional (non gravitational) interactions between the DE component and DM without conflict with the experimental data. Besides, in references \cite{pavon1} and \cite{pavon2} it is shown that interacting DE models are well consistent with current observational bounds.\footnotemark\footnotetext{It should be pointed out, however, that when the stability of DE potentials in quintessence models is considered, the coupling dark matter-dark energy is troublesome \cite{doram}. Other arguments based on oscillations or exponential blowup of the matter power spectrum that is inconsistent with observations, rule out tight coupling \cite{tegmark}.} Since there are suggestive arguments showing that observational bounds on the "fifth" force do not necessarily imply decoupling of baryons from the dark energy\cite{pavonx}, then baryons might be considered also as part of the background DM that is interacting with the quintessence field. 

Therefore, as it is done in reference \cite{pavonx}, we may consider a universal coupling of the quintessence field to all sorts of matter (radiation is excluded). Since the arguments given in the appendix of reference \cite{pavonx} to explain this possibility are also applicable in the cases of interest in the present study, we refer the interested reader to that reference to look for the details. However we want to mention the basic arguments given therein: A possible explanation is through the "longitudinal coupling" approach to inhomogeneous perturbations of the model. The longitudinal coupling involves energy transfer between matter and quintessence with no momentum transfer to matter, so that no anomalous acceleration arises. In consequence, this choice is not affected by observational bounds on "fifth" force exerted on the baryons. Other generalizations of the given approach could be considered that do involve an anomalous acceleration of the background due to its coupling to quintessence. However, due to the universal nature of the coupling, it could not be detected by differential acceleration experiments. Another argument given in \cite{pavonx} is that, since the coupling chosen is of phenomenological nature and its validity is restricted to cosmological scales (it depends on magnitudes that are only well defined in that setting), the form of the coupling at smaller scales remains unspecified. The requirements for the different couplings that could have a manifestation at these scales are that i) they give the same averaged coupling at cosmological scales, and ii) they meet the observational bounds from the local experiments. We complete the afore mentioned arguments, by noting that these are applicable even if the coupling is not of phenomenological origin like in the present investigation where the kind of coupling chosen is originated in a scalar-tensor theory of gravity.

Models with non-minimal interaction between the dark energy (whether scalar field modelled or not) and the background DM are appealing since the cosmic coincidence problem -why the energy densities of dark matter and of dark energy are of the same order precisely at present?- can be avoided or, at least, smoothed out\cite{zhang,wang,pavon}. In the present study we will show that these models are compelling, besides, because they can do the crossing of the phantom barrier ($\omega=-1$) with just a single scalar field, a possibility that is incompatible with models with minimal interaction between the scalar field and the background fluid. In the latter case the above mentioned crossing is possible only if two or more scalar fields are considered \cite{cross,quintom,lazkoz}. 

Many exact Fredmann-Robertson-Walker (FRW) solutions have been found in models in which the background fluid and the self-interacting scalar field DE have no other interaction than the gravitational one. However, no such gallery of solutions exists for the case where the coupling between the background and the scalar field is non-minimal. The present paper is partially aimed, precisely, at deriving classes of exact FRW solutions in models where, besides the gravitational interaction between the components of the cosmic fluid, additional non-gravitational interaction is also considered. To this end we extend a method formerly applied to derive FRW solutions in models with a scalar field minimally coupled to the background fluid (Ref. \cite{chimento}) to account for non-minimal coupling also. The classes of solutions found comprise both the phantom and the quintessence. For the input and coupling functions chosen, self-interaction potentials of the exponential and sinh like form arise. The importance of these kinds of potential is noticeable. Exponential potentials (see for instance reference \cite{copeland}) are found in higher-order\cite{ho} or higher-dimensional gravity\cite{hd}. These arise also in Kaluza-Klein and in string theories and due to non-perturbative effects (gaugino condensation \cite{gaugino}, for instance). Their role in cosmology has been investigated, for instance, in \cite{copeland,cosmo}. Sinh-like potentials have been studied, mainly, in the context of quintessence models of dark energy and as candidates for the dark matter\cite{tmatos}. To complement the study, in the final part of the paper, we modify the method of reference \cite{chimento} so that coincidence free solutions could be derived. Another goal of this paper is to study models which not only are free of the coincidence problem but, at the same time, do the crossing of the phantom barrier $\omega=-1$ at late times. We also give additional comments on the non predictive properties of scalar field cosmological models with or without energy transfer.   

The rest of the paper has been organized in the following way. In the next section II, we explain the details of the model of interacting components of the cosmic fluid we want to explore, including the field equations, etc. The model includes the possibility to deal with phantom fields by considering arbitrary sign of the scalar field kinetic energy. In Section III the method of reference \cite{chimento} (formerly used to generate exact FRW solutions in models without interaction between the components of the cosmic fluid), is generalized to account for models with interaction and with arbitrary sign of the kinetic energy of the DE. As an illustration the method is applied to the simplest case with minimal coupling. Section IV is devoted to deriving of exact solutions in models with non-minimal coupling between the DE and the background fluid. In section V we comment in detail on the physical content of solutions found in the former section. In section VI, emphasis is made in explaining how to deal with the coincidence problem in the model with interaction between the components of the cosmic fluid, subject of the present investigation. For this purpose, the method of section III is modified so that coincidence free solutions can be derived. In section VII we study a model that does the crossing of the phantom barrier. Finally, in Section VIII, we summarize the main achievements and shortcomings of the paper.

\section{The model}

We consider the following action that is inspired in a scalar-tensor theory written in the Einstein frame. The matter degrees of freedom and the scalar field are coupled through the scalar-tensor metric $\chi(\phi)^{-1}g_{ab}$ \cite{kaloper}:

\begin{eqnarray}
S=\int d^{4} x \sqrt{|g|}\{\frac{R}{2}-\frac{\epsilon}{2}(\nabla \phi)^{2}-V(\phi)\nonumber\\ + \chi^{-2}(\phi) {\cal
L}_{m}(\mu,\nabla \mu,\chi^{-1} g_{ab})\}.\label{action}
\end{eqnarray}
where $\epsilon=\pm 1$ ($\epsilon=-1$ for phantom DE, while $\epsilon=+1$ for quintessence), $V(\phi)$ is the scalar field self-interaction potential, ${\cal L}_m$ is the matter Lagrangian ($\mu$ is the collective notation for the matter degrees of freedom), and $\chi(\phi)^{-2}$ is the coupling function.

This action could be considered, instead, as an effective theory, implying additional non-gravitational interaction between the components of the cosmic fluid. When the coupling between the scalar field and the matter is minimal (no other interaction than the gravitational one) $\chi(\phi)=\chi_{0}=1$. The field equations derivable from the action (\ref{action}) are

\begin{eqnarray}
R_{\mu \nu}-\frac{1}{2}g_{\mu \nu}R=T^{(\phi)}_{\mu
\nu}+T^{(m)}_{\mu \nu},\label{fieldequation}
\end{eqnarray}
where

\begin{eqnarray}
&&T^{(m)}_{\mu \nu}=(\rho_{m}+p_{m})u_\mu u_\nu+p_{m}g_{\mu \nu},\nonumber\\
&&T^{(\phi)}_{\mu \nu}=\epsilon \nabla_\mu\phi \nabla_\nu\phi
-\frac{1}{2}g_{\mu \nu}\{\epsilon (\nabla\phi)^{2} +2
V(\phi)\},\label{tensorenergy}
\end{eqnarray}
are the stress-energy tensors for the ordinary matter degrees of freedom and for the self-interacting scalar field (dark energy) respectively. These fulfill the following conservation equation:

\be \nabla^\nu T^{(m)}_{\nu \mu}+ \nabla^\nu T^{(\phi)}_{\nu \mu}
=0,\label{conservationeq}\ee so that energy is not separately conserved by each one of the species in the cosmic mixture. Instead, the following dynamical equations hold:

\bea &&\nabla^\nu T^{(m)}_{\nu \mu}=-Q_\mu,\nonumber\\
&&\nabla^\nu T^{(\phi)}_{\nu \mu}=Q_\mu,\label{dynamicaleq}\eea where $Q_\mu$ is the interaction term. This is precisely the basic feature of interacting models: there is exchange of energy between the components of the cosmic fluid.

For FRW universe with metric

\begin{eqnarray}
ds^{2}=-dt^{2}+a^{2}[\frac{dr^{2}}{1-kr^{2}}+r^{2}d\Omega^{2}],\label{metric}
\end{eqnarray}
where $d\Omega^{2}=d\theta^{2}+sin^{2}\theta d\varphi^{2}$, and the spatial curvature $k=\pm1$ or $k=0$, the Friedmann and Raychaudhuri equations look like
\begin{eqnarray}
3H^{2}+\frac{3k}{a^{2}}=\rho_{m}+\rho_{\phi}+\Lambda,\label{friedmannequation}
\end{eqnarray} and

\begin{equation}
2\dot H-2\frac{k}{a^2}=-(p_m+\rho_m+p_\phi+\rho_\phi),
\label{feq2}
\end{equation}
respectively, where $\rho_{\phi}=\epsilon \dot{\phi}^{2}/2+V$ and $p_{\phi}=\epsilon \dot{\phi}^{2}/2-V$. In the last equation we have assumed $p_\phi=(\gamma_\phi-1)\rho_\phi$, where $\gamma_\phi$ is the scalar field barotropic parameter, which is related with the equation of state (EOS) parameter $\omega_\phi$ by the relationship $\gamma_\phi=\omega_\phi+1$.

The null component of the conservation equation for the matter degrees of freedom in equation (\ref{dynamicaleq}), can be written as
\begin{eqnarray}
\dot{\rho}_{m}+3H(\rho_{m}+p_{m})=Q,\label{densityconservation}
\end{eqnarray}
where the interaction term $Q$ is given by

\begin{eqnarray}
Q=\rho_{m}\frac{\dot{\bar{\chi}}}{\bar{\chi}}= \rho_{m}H
[a\frac{d(ln \bar{\chi})}{da}] ,\label{interactionfunction}
\end{eqnarray} and the following ``reduced" coupling function $\bar{\chi}=\chi^{(3\gamma_{m}-4)/2}$ ($0\leq\gamma_m\leq 2$ - the
matter barotropic index) has been introduced. Equation (\ref{densityconservation}) with $Q$ given by (\ref{interactionfunction}) can be readily integrated to yield:

\begin{eqnarray}
\rho_{m}=\rho_{m0}\;a^{-3\gamma_{m}}\bar{\chi},\label{density}
\end{eqnarray}
where we have considered that the ordinary matter degrees of freedom (the background) are in the form of a barotropic perfect fluid, so that $p_m=(\gamma_m-1)\rho_m$.

For the DE one has, instead (null-component of the second equation in (\ref{dynamicaleq})):
\begin{eqnarray}
\dot{\rho}_{\phi}+3H(\rho_{\phi}+p_{\phi})=-Q,\label{scalarconservation}
\end{eqnarray}
or, equivalently;

\begin{eqnarray}
\dot{\phi}[\epsilon \ddot{\phi}+ 3\epsilon H \dot{\phi}+ V']=
-Q.\label{kleingordon}
\end{eqnarray}

Equations (\ref{friedmannequation}), (\ref{interactionfunction}), (\ref{density}) and (\ref{kleingordon}) represent the basic set of equations of the model of interacting components of the cosmic fluid we are about to investigate. In what follows we shall apply a method for deriving new exact solutions to this set of equations.

\section{The Method}

In this section we will generalize the method developed in reference \cite{chimento} so that we can derive exact FRW solutions in models with interaction, such as the model detailed in the former section. To this end, lets assume that the relevant functions can be given in terms of the scale factor: $\rho_{m}=\rho_{m}(a)$, $\rho_{\phi}=\rho_{\phi}(a)$,
$\bar{\chi}=\bar{\chi}(a)$, etc. If we introduce the new time variable: $d\eta=a^{-3}dt$, then equation (\ref{kleingordon}) can be written as

\begin{eqnarray}
\frac{d}{d\eta}[\frac{\epsilon}{2}(\frac{d\phi}{d\eta})^{2}+ F]=
6\frac{F}{a}\frac{da}{d\eta}-
\rho_{m0}a^{3(2-\gamma_{m})}\frac{d\bar{\chi}}{d\eta},\label{kleingordon2}
\end{eqnarray}
where we have introduced the input function $F=F(a)$. This function is chosen in such a way that the self-interaction
potential $V$ can be rewritten as function of the scale factor in the following form\cite{chimento}:

\begin{eqnarray}
V(a)=\frac{F(a)}{a^{6}}.\label{VFunction}
\end{eqnarray}

Equation (\ref{kleingordon2}) can be readily integrated to yield

\begin{eqnarray}
a^{6}\rho_{\phi}=\int da[6\frac{F}{a}-
\rho_{m0}a^{3(2-\gamma_{m})}\frac{d\bar{\chi}}{da}] + C,\label{kleingordon3}
\end{eqnarray}
where $C$ is an arbitrary integration constant. If we integrate by parts the second term in the right hand side (RHS) of (\ref{kleingordon3}), then we are led with the following equation:

\bea a^{6}(\rho_{\phi}+\rho_{m})=&&3 \int
\frac{da}{a}[2F+\nonumber\\
&&(2-\gamma_{m})\rho_{m0}a^{3(2-\gamma_{m})}\bar{\chi}]+C.\label{kleingordon4}
\eea

We now introduce the functions \bea G(a)\equiv 3H^{2}=\rho_{m}+\rho_{\phi}+\Lambda-3k/a^2,\label{Gfunction} \eea and

\bea L(a)\equiv \epsilon (\rho_{\phi}-V)= \epsilon(\rho_{\phi}-\frac{F}{a^{6}}),\label{Lfunction} \eea that will be useful in what follows. Both functions $G(a)$ and $L(a)=\dot{\phi}^{2}/2$ are always non-negative: $G(a)\geq 0$, $L(a)\geq 0$. The cosmological constant $\Lambda$ can be absorbed into the self-interaction potential $V(\phi)$ so we can set $\Lambda=0$ without loss of generality. 

In what follows, for sake of simplicity and unless the contrary is specified, we choose the spatial curvature $k=0$, i. e., we will explore flat FRW cosmologies. Another magnitude of interest, that will be useful in the future, is the DE barotropic parameter:

\bea \gamma_{\phi}=\frac{2a^{6}L(a)}{a^{6}L(a)+\epsilon F}.\label{eos} \eea

After considering equations (\ref{kleingordon3}) and (\ref{kleingordon4}), the functions $G(a)$ and $L(a)$ can be written in the following form: 

\bea
G(a)=&&\frac{3}{a^{6}} \int \frac{da}{a}[2F+\nonumber\\
&&(2-\gamma_{m})\rho_{m0}a^{3(2-\gamma_{m})}\bar{\chi}] + C/a^6,\label{GFunction2} \eea and

\begin{equation}
L(a)=\epsilon G(a)-\frac{\epsilon}{a^{6}}(F+\rho_{m0}a^{3(2-\gamma_{m})}\bar{\chi}),\label{LFunction2}
\end{equation}
respectively.

Exact solutions can be found in the form of quadratures\cite{chimento}:

\begin{eqnarray}
\Delta t=\pm\sqrt{3}\int\frac{da}{a\sqrt{G(a)}},\label{deltaT}
\end{eqnarray}
or

\begin{eqnarray}
\Delta \eta=
\pm\sqrt{3}\int\frac{da}{a\sqrt{a^{6}G(a)}},\label{deltaEta}
\end{eqnarray}
and

\begin{eqnarray}
\Delta
\phi=\pm\sqrt{6}\int\frac{da}{a}\sqrt{\frac{L(a)}{G(a)}}.\label{deltaFi}
\end{eqnarray}

In equations (\ref{deltaT}), (\ref{deltaEta}) the ``$\pm$" sign in the RHS means both time directions and, since Einstein's equations (equations (\ref{friedmannequation}), (\ref{densityconservation}), (\ref{interactionfunction}) and (\ref{scalarconservation})) are invariant under time inversion then, in what follows, we choose the branch with the ``+" sign in equation (\ref{deltaT}) (or (\ref{deltaEta})).

Once the input function $F=F(a)$ and the coupling function $\bar{\chi}=\bar{\chi}(a)$ are given, we can find $G=G(a)$ and $L=L(a)$ through equations (\ref{GFunction2}) and (\ref{LFunction2}) respectively. Then, by use of equations (\ref{deltaT}) (or (\ref{deltaEta})) and (\ref{deltaFi}), one is able to find $t=t(a)$ (or $\eta=\eta(a)$) and $\phi=\phi(a)$ by direct integration and, by inversion, $a=a(t)$ (or $a=a(\eta)$) and $\phi=\phi(t)$ (or $\phi=\phi(\eta)$) respectively.

Let us illustrate how the method operates through the study of the simplest example: a model without interaction $\bar{\chi}=\bar{\chi}_{0}=1$ ($Q=0$). I. e., the DE and the background fluid evolve independently. In this case (see equations (\ref{GFunction2}) and (\ref{LFunction2})):

\begin{eqnarray}
G(a)=\rho_{m0}a^{-3\gamma_{m}}+\frac{6}{a^6} \{\int \frac{da}{a}F(a) +C/6\},\label{Gcase1}
\end{eqnarray}
and

\begin{eqnarray}
L(a)=\frac{\epsilon}{a^6}\{6 \int \frac{da}{a}F(a) -F(a) +C\}.\label{Lcase1}
\end{eqnarray}
For simplicity we set $C=0$. To chose the input function we assume, additionally\footnotemark\footnotetext{Other assumptions to fix the input function can also be made. In principle the input function is arbitrary, the only requirement being simplicity of the integrals to be taken.}

\begin{eqnarray}
\int \frac{da}{a}F(a)=\frac{1}{s}F(a),\label{Equ314}
\end{eqnarray}
where $s$ is some arbitrary constant parameter. The equation (\ref{Equ314}), in particular, is fulfilled if $F(a)= B\;a^{s}$, where $B$ is a constant parameter. Therefore

\begin{eqnarray}
&&G(a)=\rho_{m0}\;a^{-3\gamma_{m}}+\frac{6B}{s}a^{s-6},\nonumber\\ 
&&L(a)=\epsilon(\frac{6-s}{s})Ba^{s-6}.\label{Equ316}
\end{eqnarray}

Taking into account all of the above, the equation (\ref{eos}) yields to:

\be \gamma_\phi=\frac{6-s}{3}.\label{caseminimal}\ee Since, for quintessence $0<\gamma_\phi<2/3$, then $4<s<6$. For $s>6$ ($\gamma_\phi<0$) the DE is a phantom field instead. Let us investigate the most general situation when $\gamma_m\neq (6-s)/3=\gamma_\phi$. In this case the solutions are given by the following expressions:

\begin{eqnarray}
\Delta t=(\frac{2}{3\gamma_m})\sqrt{\frac{3}{\rho_{m0}}}\;\; a^\frac{3\gamma_m}{2}\;
\;_2F_1[\frac{3\gamma_m}{2(3\gamma_m+s-6)},\frac{1}{2},\nonumber\\
1+\frac{3\gamma_m}{2(3\gamma_m+s-6)},-(\frac{6B}{s\rho_{m0}})^\frac{3\gamma_m+s-6}{3\gamma_m}a^{3\gamma_m+s-6}], \label{Equ320}
\end{eqnarray} where $\;_2F_1$ is the hypergeometric function, and

\begin{eqnarray} \Delta \phi=\frac{\sqrt{2\epsilon/n}}{k} arc\sinh [\sqrt{\frac{6B}{s\rho_{m0}}}a^{-k}],\nonumber\\
k=-\frac{s-6+3\gamma_m}{2},\;\;n=\frac{2}{6-s},\label{Equ321}
\end{eqnarray}
so, by inversion of (\ref{Equ321}), one gets $a=a(\phi)$ and, considering equation (\ref{VFunction}), one finds the form of the potential:

\begin{eqnarray}
V(\phi)=V_0\sinh^{2q}[\lambda\Delta\phi], \label{Equ322}
\end{eqnarray}
where $V_0=B(s\rho_{m0}/6B)^q$, $q=1/kn$ and $\lambda=\pm k/\sqrt{2\epsilon/n}$. Note that the constant parameter $q$ can be written also in the form $q=-\gamma_\phi/(\gamma_m-\gamma_\phi)$.
Then, if $\phi$ were a phantom ($\gamma_\phi<0\Rightarrow s>6$), $q$ were always positive.

As explained in \cite{tmatos}, this potential is a good quintessencial candidate to be the missing energy in the universe. Since it behaves like an inverse power-law potential at early times, then this allows to avoid the fine tuning problem. The parameters $V_0$, $\lambda$ and $q$, can be determined uniquely by the measured values for the equation of state and of the amount of vacuum energy necessary to obtain a tracker solution \cite{tmatos}.

\section{Generating solutions in non-minimal coupling cases}

In this section we will apply the method explained in the former section, to generate solutions in models where the DE (either quintessence or phantom) and the background fluid share additional non-gravitational interaction through the non-minimal coupling given by the coupling function $\chi^{-2}$ (or, alternatively, the ``reduced'' coupling function $\bar\chi$) in the action (\ref{action}). Through this section, the constant parameters $n$, $k$, etc., are different, in general, from those in the former section.

As before, for the sake of simplicity, we consider $C=0$. In this case (see equations (\ref{GFunction2}) and (\ref{LFunction2})):

\begin{eqnarray}
&&G(a)=\frac{3}{a^{6}}\int\frac{da}{a}[2F+(2-\gamma_{m})\rho_{m0} a^{3(2-\gamma_{m})}\bar\chi],\nonumber\\
&&L(a)=\epsilon\{G(a)-\frac{F}{a^6}-\rho_{m0}a^{-3\gamma_{m}}\bar\chi\},\label{Equ39}
\end{eqnarray}
and
\begin{eqnarray}
\frac{L(a)}{G(a)}=\epsilon\{1-\frac{F+\rho_{m0} a^{3(2-\gamma_{m})}\bar\chi}{a^{6}G(a)}\}.\label{Equ40}
\end{eqnarray}

Now we are in position to introduce different input $F(a)$ and coupling functions $\bar\chi (a)$ to generate solutions. However, to illustrate the possibilities offered by the method, it will be enough to choose just one given input function and a couple of coupling functions.

\subsection{$F=B\; a^s$, $\bar\chi=\bar\chi_0\;a^{3\gamma_m-n}$}

After this choice of the input functions $F$ and $\bar\chi$, straightforward integration in the first equation in (\ref{Equ39}) yields to:

\be G(a)=\frac{6B}{s}\;a^{s-6}+\frac{A}{6-n}\;a^{-n},\label{Equ41}\ee where $n\neq 6$ and $A=3\bar\chi_0 \rho_{m0}(2-\gamma_m)$. Meanwhile, Eq. (\ref{Equ40}) can be written as follows:

\bea &&\frac{L}{G}=\epsilon [\frac{(6-s)B\;a^{s-6}+\alpha\;a^{-n}}{6B\;a^{s-6}+\frac{s A}{6-n}\;a^{-n}}],\nonumber\\
&&\alpha=\frac{s(n-3\gamma_m)A}{3(6-n)(2-\gamma_m)},\label{Equ42}\eea
and Eq(\ref{eos}) for the scalar field barotropic parameter:

\be \gamma_\phi=2[\frac{(6-s)B\;a^{s-6}+\alpha\;a^{-n}}{6B\;a^{s-6}+\alpha\;a^{-n}}].\label{eosA}\ee

We point out that, since $G(a)\geq 0$, then the constant parameter $n$ is restricted to be $n<6$.

\subsubsection{General case with $s\neq 6$}

\bigskip

(i) $\alpha=0\Rightarrow n=3\gamma_m$

\bigskip

In this case there is no additional (non-gravitational) interaction between the scalar field and the background fluid: $\bar\chi=\bar\chi_0$, i.e., this is the simplest situation where minimal coupling between the dark energy and the background fluid is considered. According to equation (\ref{eosA}) the DE barotropic parameter is related with the constant parameter $s$ through: $3\gamma_\phi=6-s$. In consequence, for quintessence ($0\leq\gamma_\phi<2/3$) $4<s\leq 6$, meanwhile, for the phantom ($\gamma_\phi< 0$) $s>6$.

Obviously, if one correctly arranges the constants, this case is the same one considered, as a working example, in the last part of the former section. To simplify writing of the solutions, lets introduce the constant $k=\frac{n}{6-s}-1$. The integral in (\ref{deltaFi}) is now easily taken to yield, for $k\neq 0$:

\be \Delta\phi=\pm \frac{2\sqrt{\epsilon
(6-s)}}{k(6-s)}\;arc\sinh[\sqrt\frac{6(6-n)B}{s A}\;Y^{-k/2}],\label{deltaFi1}\ee or, after inverting it to get $a=a(\phi)$:

\bea &&V(\phi)=V_0\;\sinh^{-2/k}[\lambda\Delta\phi],\nonumber\\
&&V_0=B[\frac{6(6-n)B}{s
A}]^{1/k},\;\;\lambda=\pm\frac{k(6-s)}{2\sqrt{\epsilon(6-s)}}.\label{VAi}\eea

Equation (\ref{deltaT}) can now be rewritten as follows:

\be \Delta
t=\pm\frac{1}{n}\sqrt\frac{2s}{B}\int\frac{dX}{\sqrt{X^{2p}+\frac{s
A}{6(6-n)B}}},\label{deltaTA1i}\ee where we have introduced the new variable $X=a^{n/2}$ and the new constant parameter $p=k/(k+1)$. The integral (\ref{deltaTA1i}) yields to:

\bea \Delta t=\pm\frac{2}{n}\sqrt\frac{3(6-n)}{A}\;a^{n/2}
\;_2F_1[\frac{1}{2p},1,\nonumber\\
1+\frac{1}{2p},\frac{6p(n-6)B}{s A}\;a^n],\label{deltaTA1i'}\eea where, as before, $_2F_1$ is the hypergeometric function.

\bigskip

(ii) $\alpha\neq 0$, $k=0$ ($n=6-s$)

\bigskip

The "reduced" coupling function $\bar\chi$ is now of the form $\bar\chi=\bar\chi_0\; a^{3\gamma_m+s-6}$. Since, in this case (see equations (\ref{Equ41}) and (\ref{Equ42})),

\bea G(a)=\frac{6B+A}{s}\; a^{s-6},\nonumber\\
\frac{L}{G}=\epsilon q,\;\;
q=\frac{6-s+\alpha/B}{6+A/B},\label{GLAii}\eea the evolution of the scale factor evolution in terms of cosmic time is given by

\bea &&a(t)=a_0 \Delta t^{2/(6-s)},\nonumber\\
&&a_0=[(\frac{6-s}{2})^2(\frac{6B+A}{3s})]^{1/(6-s)},\eea
meanwhile the self-interaction potential is given by:

\be V(\phi)=V_0\;
e^{-\lambda\Delta\phi},\;\;\lambda=\pm\frac{6-s}{\sqrt{6\epsilon
q}},\label{VAii}\ee where $V_0=B$. Note that there is scaling of the form $\rho_m/\rho_\phi=const$. In the present case, the barotropic parameter $\gamma_\phi$ is given by

\be \gamma_\phi=2[1-\frac{s B}{\alpha+6B}]=const.\label{eosAii}\ee

\subsubsection{Particular case with $s=6$}

If in equations (\ref{Equ41}-\ref{eosA}) one considers $s=6$, i.e., $V=V_0=B$ then, one is faced with a situation where the sources of gravity are a scalar field without self-interaction potential (only a kinetic energy term present), and the background perfect fluid (dark matter, for instance), both ``living'' in a de Sitter background space-time. It is fixed by the effective cosmological constant $\Lambda=B$. The following equations hold:

\bea &&G(a)=B+A\;a^{-n}/(6-n),\nonumber\\
&&\frac{L}{G}=\epsilon \frac{\alpha'\; a^{-n}}{6B+\frac{6
A}{6-n}a^{-n}},\;\;\alpha'=\frac{2(n-3\gamma_m)A}
{(6-n)(2-\gamma_m)}.\label{GL}\eea

As customary, by taking the integral in equation (\ref{deltaT}), one obtains the evolution of the scale factor in cosmic time:

\be a(t)=a_0\;\sinh^{2/n}[\frac{n}{2}\sqrt\frac{B}{3}\Delta t],\;\; a_0=[\frac{A}{(6-n)B}]^{1/n},\label{scalefA1}\ee
meanwhile, the evolution of the scalar field is given by:

\bea \Delta\phi=\pm\frac{2}{n}\sqrt\frac{\epsilon\alpha'(6-n)}{A} \times\nonumber\\
arc\sinh[\sqrt\frac{A}{(6-n)B}\;a^{-n/2}].\label{deltaFiA1}\eea
The scalar field energy density is contributed by the effective cosmological constant\footnotemark\footnotetext{As already said, this effective cosmological constant can be interpreted, alternatively, as a background vacuum fluid.} and the scalar field kinetic energy density:

\be \rho_\phi=\frac{\alpha'}{6}a^{-n}+B.\label{rhophi}\ee Notice that, the quintessence solution ($\epsilon=+1$) arises whenever $\alpha'>0\Rightarrow 0<n<3\gamma_m$ (recall that $n<6$), meanwhile the phantom behavior is displayed once $3\gamma_m<n<6$.\footnotemark\footnotetext{The possibility $n<0$ is ruled out since, in this case, according to (\ref{rhophi}), the negative energy component of the phantom field increases with the expansion.} However, since $\alpha'<0$ for the phantom scalar, there is a time $t_c$, such that $a^n(t_c)=a_c^n=-\alpha'/6B\Rightarrow \rho_\phi(t_c)=0$. For earlier times $t<t_c$, the energy density of the phantom field is negative definite. This fact alone, rules out the possibility to describe phantom behavior with the help of the present solution, unless the free parameters are chosen in such a way that $t_c$ is close enough to the Planck time scale. In this case one might argue that the classical theory of gravity is unable to give an appropriated (perhaps semiclassical or quantum) description of the cosmic evolution.

As it will be discussed in the next section, this solution represents an example of a model with transition from decelerated into accelerated expansion, where the accelerated phase is driven by the combined effect of the kinetic energy density of the scalar field and of the cosmological constant.

In the next subsection we will investigate cases with the same input function $F(a)=B\;a^s$ and a different coupling function $\bar\chi(a)$.

\subsection{$F=B\; a^s$, $\bar\chi=\bar\chi_0
\;a^{3\gamma_m}[a^{-n}+a^{s-6}]$}

We write the relevant functions in this case:

\bea G(a)=\frac{6B'}{s}\; a^{s-6}+\frac{A}{6-n}\;
a^{-n},\nonumber\\
B'=B+A/6,\;\;A=3(2-\gamma_m)\rho_{m0}\bar\chi_0,\label{GB}\eea and
\bea &&\frac{L}{G}=\epsilon
[\frac{\alpha\;a^{s-6}+\beta\;a^{-n}}{\frac{6B'}{s}
\;a^{s-6}+\frac{A}{6-n}\;a^{-n}}],\nonumber\\
&&\alpha=\frac{(6-s)B'}{s}-\frac{\gamma_m
A}{6(2-\gamma_m)},\nonumber\\
&&\beta=\frac{(n-3\gamma_m)A} {3(6-n)(2-\gamma_m)},\label{LGB}\eea
and the barotropic index for the scalar fluid:

\be \gamma_\phi=2\frac{\alpha\;a^{s-6}+\beta\;a^{-n}}{(\alpha+B)\;a^{s-6}+\beta\;a^{-n}}.\label{eosB}\ee

We should note that positivity of the energy density (non-negativity of the function $G(a)$), restricts the parameter $n$ to be $n<6$. Classes of exact solutions can be easily found for particular relationships among the free parameters.

\subsubsection{General case with $s\neq 6$}

As before, since the input function $F(a)=B\;a^s\Rightarrow V=B\;a^{s-6}$. It is also useful to introduce the constant
$k=n/(6-s)-1$. For $n>6-s$, $k>0$ meanwhile, for $n<6-s$, $k<0$.
Straightforward integration in (\ref{deltaT}) yields to:

\bea \Delta
t=\pm\frac{\sqrt{3(6-n)/A}}{2(s-6)(k+1)}\;Y^{-\frac{k+1}{2}}\times\nonumber\\
\;_2F_1[\frac{k+1}{2k},\frac{1}{2},\frac{3k+1}{2k},\frac{6(n-6)B'}{s
A}\;Y^{-k}].\label{deltaTsneq6}\eea

Let us try, as before, particular cases where the integral in Eq. (\ref{deltaFi}) is easily taken.

\bigskip

(i) $\alpha=0$

\bigskip

This choice implies the following relationship involving the constant parameters $A$, $B$, $s$ and $\gamma_m$:

\be
B=[\frac{s-6+3\gamma_m}{3(2-\gamma_m)(6-s)}]A.\label{constantsB}\ee
Positivity of the constant $B$ leads to the following restriction on the parameter $s$: $3(2-\gamma_m)<s<6$. After the choice of the relationship between the constant parameters made, if one integrates (\ref{deltaFi}) to find $\Delta\phi$ and then inverts to get $V=V(\phi)$, one obtains:

\bea
&&V(\phi)=V_0\sinh^{2/k}[\lambda\Delta\phi],\nonumber\\&&V_0=B[\frac{6(6-n)B'}{s
A}]^{1/k},\;\;\lambda=\pm\sqrt\frac{(s-6)^2A}{6\epsilon\beta(6-n)}.\label{VBi}\eea

Since $n<6$, and since $\lambda$ should be real then, for quintessence ($\epsilon=+1$) $\beta>0\Rightarrow 3\gamma_m<n<6$, meanwhile for phantom ($\epsilon=-1$) $\beta<0\Rightarrow n<3\gamma_m$.

\bigskip

(ii) $\beta=0\Rightarrow n=3\gamma_m$

\bigskip

The coupling function is given now by

\be
\bar\chi=\bar\chi_0(1+a^{3\gamma_m+s-6}).\label{couplingBii}\ee
Integration of Eq. (\ref{deltaFi}) yields (recall that Eq. (\ref{deltaTsneq6}), representing part of the solution, is yet valid):

\be \Delta\phi=\pm\frac{2\sqrt{\epsilon\alpha s/B'}}{k(s-6)}\;
arc\sinh[\sqrt\frac{6(6-n)B'}{s A}\;Y^{-k/2}],\label{deltaFisol}\ee or, after inversion:

\bea &&V(\phi)=V_0\;\sinh^{-2/k}[\lambda\Delta\phi],\nonumber\\&&V_0=B[\frac{6(6-n)B'}{s
A}]^{1/k},\;\;\lambda=\pm\frac{k(s-6)}{2\sqrt{\epsilon\alpha s/B'}}.\label{VFi}\eea We see that, for a phantom field
($\epsilon=-1$), $\alpha<0$ and viceversa.

\subsubsection{Particular case with $s=6\Rightarrow V=V_0=B$}

We are faced with a scalar field without self-interaction potential. The coupling function looks like:

\be \bar\chi=\bar\chi_0
a^{3\gamma_m}(1+a^{-n}),\label{couplingf}\ee Since $\rho_m=\rho_{m,0}\;a^{-3\gamma_m}\;\bar\chi$, the background
fluid can be viewed as a mixture of a vacuum fluid with a constant energy density plus a perfect fluid with barotropic index $n/3$.

The other relevant functions are given as follows:

\bea G(a)=B'+\frac{A}{6-n}\;a^{-n},\nonumber\\\frac{L}{G}=\epsilon\frac{\alpha+
\beta\;a^{-n}}{B'+\frac{A}{6-n}\;a^{-n}},\label{relevantf}\eea
where now the constant $\alpha$ is always negative since:

\be \alpha=-\frac{\gamma_m A}{6(2-\gamma_m)},\label{alpha}\ee
while $A>0$ and $0\leq\gamma_m\leq 2$. If we integrate (\ref{deltaT}) and then invert, we obtain the evolution of the
scale factor in cosmic time:

\bea &&a(t)=a_0\;\sinh^{2/n}[\frac{n}{2}\sqrt\frac{B'}{3}\Delta
t],\nonumber\\
&&a_0=[\frac{A}{(6-n)B'}]^{1/n}.\label{at}\eea At the same time, by considering equation (\ref{relevantf}), straightforward integration of (\ref{deltaFi}) leads to:

\bea &&\Delta\phi=\pm\frac{\sqrt{6\epsilon\alpha/B'}}{n}
\{\ln[2x+b+c+2\sqrt{(x+b)(x+c)}]\nonumber\\
&&-\sqrt\frac{b}{c}\ln[\frac{2\sqrt{cb(x+b)(x+c)}+cx+b(x+2c)}
{\sqrt{cb^3}x}]\},\nonumber\\
&&x=a^n,\;\;b=\frac{\alpha}{\beta},\;\;c=\frac{A}{(6-n)B'}.
\label{dphi}\eea
This last integral is easier to take and, in correspondence, a simpler relationship between $\phi$ and the scale factor $a$ is obtained, when one considers the particular case $\beta=0$ ($n=3\gamma_m$). In this case the coupling function takes the form: $\bar\chi=\bar\chi_0(1+a^{3\gamma_m})$. The scale factor evolves according to Eq. (\ref{at}) as before, however, the integral in (\ref{deltaFi}) yields to:

\be \Delta\phi=\pm\frac{2\sqrt{6\epsilon\alpha/B'}}{n}\;arc\sinh[\sqrt\frac{(6-n)B'}{A}\;a^{n/2}].\label{dphi'}\ee In
this case only phantom fields can be accommodated since $\alpha$ is negative.

Although we have not exhausted all the possibilities offered by the method to generate exact solutions, we think the
aforementioned solutions illustrate quite well its power. In particular, one can try with other input functions $F(a)$ and coupling functions $\bar\chi(a)$, etc. In the next section we will comment in detail the solutions found in a more physical context.

\section{The physical significance of the mathematical solutions}

For the subsequent discussion it will be useful to give the expressions that determine several important parameters of physical (and observational) importance like the Hubble parameter $H$ or its derivative in respect to the cosmic time $\dot H$ (these are useful to determine whether the expansion is accelerated or decelerated), the dimensionless energy density parameter for the scalar field $\Omega_\phi=\rho_\phi/3H^2$, etc.\footnotemark\footnotetext{Other parameters of physical importance, as for instance, the barotropic parameter of the scalar field fluid $\gamma_\phi$, have been defined in precedent sections (see equations (\ref{eosA}), (\ref{eosAii}), (\ref{eosB}), etc.).}Below we collect the general expressions for them.

\bea &&H=\pm\sqrt{G(a)/3},\;\;\Omega_\phi=1 -\rho_{m0}a^{-3\gamma_m}\frac{\bar\chi}{G},\nonumber\\ &&\dot H=G\{(\gamma_m-\gamma_\phi)\Omega_\phi-\gamma_m\}/2,
\label{physparameters}\eea so, in particular, acceleration of the expansion:

\be \frac{\ddot a}{a}=\dot H+H^2=G\{3(\gamma_m-\gamma_\phi)\Omega_\phi-3\gamma_m+2\}/6,
\label{acceleration}\ee is allowed whenever

\be \ddot a>0\;\Rightarrow\;\; 3(\gamma_m-\gamma_\phi)\Omega_\phi>3\gamma_m-2.\label{accelerationcondition}\ee

In this section we will show that it always possible to find a region in (free) parameter space where the given solutions are adequate to describe correctly the present cosmological paradigm. We have organized the discussion of the solutions previously found, in the same way as these solutions were organized in Section IV. This means that we first discuss the solution of subsection A.1 (i), then A.1 (ii), etc.

\subsection{$F=B\; a^s$, $\bar\chi=\bar\chi_0\;a^{3\gamma_m-n}$}

\subsubsection{General case with $s\neq 6$}

\bigskip

(i) $\alpha=0\Rightarrow n=3\gamma_m$

\bigskip

In this case, as already said, there is no additional interaction between the DE and the DM ($\bar\chi=\bar\chi_0$). The scalar field (DE) barotropic parameter is a constant:
$\gamma_\phi=(6-s)/3$. The expression of the DE dimensionless energy density is given by:

\be \Omega_\phi=\frac{\epsilon+
\beta\;a^{3(\gamma_m-\gamma_\phi)}}{1+
\beta\;a^{3(\gamma_m-\gamma_\phi)}},\ee where $\epsilon=1-3(2-\gamma_m)\rho_{m0}\bar\chi_0/A$, is the fraction of DE at the beginning of the expansion ($\epsilon\ll 1$), and $\beta=6(2-\gamma_m)B/(2-\gamma_\phi)A$. Recall that the solution comprises both quintessence ($0\leq \gamma_\phi<2/3$) and phantom ($\gamma_\phi<0$) behavior. Notice that, at late times, the dynamics of the expansion is completely dominated by the scalar field: $\Omega_\phi\rightarrow 1$. Accelerated expansion is allowed once

\be \Omega_\phi>\frac{3\gamma_m-2}{3\gamma_m+s-6},\ee or, if one considers that the DM is dust ($\gamma_m=1$): $3\Omega_\phi>1/(s-3)$. Accordingly, this solution allows for transition from decelerated into accelerated expansion, as required by the cosmological paradigm that dominates at present. Since the DE density is given by:

\be \rho_\phi=a^{-3\gamma_m}\{\frac{A}{6-n}-\rho_{m0}\bar\chi_0 +\frac{6B}{s}a^{3(\gamma_m-\gamma_\phi)}\},\ee then, for phantom DE ($\gamma_\phi<0\Rightarrow s>6$), the energy density is high at early times and is a decreasing function, until a moment when it begins to grow. In particular, at late times, the energy density is a manifestly increasing function of the cosmic time, and it becomes infinitely large in a finite amount of proper time, signaling at a Big-Rip type of singularity in the future of the cosmic evolution. This is a generic feature of (non-interacting)
models of phantom energy.

\bigskip

(ii) $\alpha\neq 0$, $k=0$ ($n=6-s$)

\bigskip

In this case, the coupling function is of the following form: $\bar\chi=\bar\chi_0 a^{3\gamma_m+s-6}$, while the dimensionless energy density parameter for the DE is a constant. It is given by:

\be \Omega_\phi=\frac{6B+\alpha}{6B+A}.\ee The DE barotropic parameter is also a constant:
$\gamma_\phi=2((6-s)B+\alpha)/(6B+\alpha)$. In fact, since $\Omega_\phi=const,\Rightarrow\Omega_m=const,\Rightarrow
\Omega_m/\Omega_\phi=const$, then if the DE is currently dominating the cosmic evolution, it should dominate the entire cosmic history, contrary to the evidence from nucleosynthesis about matter dominance in the past. In other words, in the present scenario there is no room for transition from matter dominance in the past, to dark energy dominance at present, but it is possible to find a bounded region of values for the $(A, B, s, \alpha)$ parameters, where this solution describes the accelerated expansion at late time. As easily seen, this solution, with a single exponential potential, is not appropriated for the description of the presently accepted cosmological paradigm. This situation is similar when we have minimal coupling with a scalar field with single exponential self-interaction potential.

\subsubsection{Particular case with $s=6$}

This solution is very peculiar since the sources of gravity are a scalar field without self-interaction potential (only a scalar field kinetic energy term is present) and the background perfect fluid, both ``living'' in a de Sitter background space-time. It is fixed by the effective cosmological constant $\Lambda=B$. The DE barotropic parameter is given by:

\be \gamma_\phi=\frac{2\alpha'}{6B a^n+\alpha'},\label{DEEOS}\ee
meanwhile, the energy density parameter:

\be \Omega_\phi=\frac{Ba^n+\alpha'/6}{Ba^n+A/(6-n)}.\ee Worth noting that the DE barotropic parameter is dynamical, i.e., it evolves during the cosmic evolution. It is seen that smooth transition from stiff matter in the past, into vacuum fluid in the future, is obtained. The condition to get accelerated expansion is translated into the following requirement:

\be
a^n>(\frac{3\gamma_m-2}{6-n})\frac{A}{2B}+(1-\gamma_m)\frac{\alpha'}{4B}.\ee
Notice that, in the present case, transition from decelerated into accelerated expansion is allowed. This means that we are faced with a model where the acceleration of the cosmic expansion is driven by the combined action of the vacuum (effective cosmological constant) energy density $\Lambda_0=B$, and the kinetic energy density of the scalar field. If we neglect the kinetic energy of the scalar field: $\alpha'=0\Rightarrow n=3\gamma_m$, then we recover the standard $\Lambda$ cold dark matter ($\Lambda$CDM) scenario, without interaction between the DM and the background vacuum fluid. This fact provides the following interpretation of this particular solution: the interaction between the scalar field and the background fluid translates into a change only in the kinetic energy of the scalar field, its potential energy is preserved during the exchange.

As explained in the former section, if the free parameters of this solution are chosen in such a way that $t_c\Rightarrow \rho_\phi(t_c)=0$ is close enough to the Planck time scale, the solution could be able to describe phantom behavior also. However, unlike a large body of phantom models of DE, the cosmic evolution proceeds without big-rip event in the future. This fact is easily explained with the help of equation (\ref{rhophi}) for the energy
density of the phantom field. Actually, notice that the energy density decreases in cosmic time $t$, contrary to the requirement of an increasing energy density to reach a big-rip singularity in a finite amount of cosmic time into the future.

\subsection{$F=B\; a^s$, $\bar\chi=\bar\chi_0
\;a^{3\gamma_m}[a^{-n}+a^{s-6}]$}

\subsubsection{General case with $s\neq 6$}

\bigskip

(i) $\alpha=0$

\bigskip

This solution is adequate to describe correctly the present cosmological paradigm, through an appropriate choice of the free parameters. The barotropic parameter $\gamma_\phi$ of the DE, as a function of the scale factor, is given by:

\be \gamma_\phi (a)=\frac{2\beta}{\beta+B\;a^{n+-6}},\ee so that, for $n+s-6>0$, and $\beta>0$, the dark energy evolves from being stiff matter in the past, to being vacuum energy in the future. The same asymptotic behavior holds for the case when $\beta<0$, but the DE behaves like a phantom fluid in the meanwhile.

A very appealing feature of the present solution, is that the cosmic coincidence does not even arise. In fact, choose
$k>0\Rightarrow n>6-s$. Lets impose just one more condition on the constant parameters at the beginning of the expansion $a(0)=0$:

\be \Omega_\phi
(0)=0\;\Rightarrow\;\rho_{m0}\bar\chi_0=\frac{A}{6-n},\ee that is compatible with nucleosynthesis constrains on the amount of dark energy at such early times. This restriction implies that, in the distant future ($a(\infty)=\infty$):

\be \Omega_\phi (\infty)=\frac{(6-n)B-(n+s-6)A}{(6-n)(B+A/6)},\ee
where, since $0\leq\Omega_\phi\leq 1$; then:

\be B>\frac{n+s-6}{6-n}(A/6).\label{constrel}\ee Therefore, as the cosmic expansion proceeds, the ratio between the background of dark matter and the dark energy
$r=\Omega_m/\Omega_\phi=\rho_m/\rho_\phi$ tends to a constant value $r_0=(1-\Omega_\phi (\infty))/\Omega_\phi (\infty)\lesssim 1$. If the free parameters are correctly chosen, it may happen that a state with almost constant ratio $r\approx r_0$ has been already reached and, the future of the cosmic evolution is just to asymptotically approach $r_0$. If this is indeed the case, then the ``cosmic coincidence'' is not such a coincidence, but it is the result of the predictable destiny of the cosmic evolution.

\bigskip

(ii) $\beta=0\Rightarrow n=3\gamma_m$

\bigskip

Here we have a similar situation, with the distinction that the dark energy barotropic parameter is always a constant:

\be \gamma_\phi =\frac{2\alpha}{\alpha+B},\ee where $\alpha>0$ is for standard DE, meanwhile $\alpha<0$ is for the phantom. If one chooses the constant $A=3(2-\gamma_m)\rho_{m0}\bar\chi_0$, then the DE dimensionless energy density parameter $\Omega_\phi$ evolves from a zero value in the distant past into a value:

\be \Omega_\phi
(\infty)=\frac{6(2-\gamma_m)B-(3\gamma_m+s-6)A/3}{(2-\gamma_m)(6B+A)}<\sim 1\ee through an appropriate arrange of the free parameters. This means that the coincidence problem can be evaded also in this case. This is not surprising, because, as already mentioned, models where there is exchange of energy and/or moment between the components of the cosmic mixture are adequate to address this issue.

\subsubsection{Particular case with $s=6\Rightarrow V=V_0=B$}

Finally we have the situation where one is faced with a scalar field without self-interaction potential. As already said, in this case only phantom fields can be accommodated since $\alpha$ is negative.

\section{Avoiding the Cosmic Coincidence Problem}\label{coincidence}

Since models with interaction between the components of the cosmic fluid are appropriated for handling of the cosmic coincidence problem (CCP): why the density parameters of the DM and of the DE are simultaneously non-vanishing precisely at present? in order to complement the present analysis, we develop in this section a modification of the method previously used to derive FRW solutions, so that we will be able to derive FRW solutions that are free of the cosmic coincide.

Here we follow a procedure similar to that of reference \cite{pavonx}. Let us investigate first, under which conditions the aforementioned question does not arise in the situations of interest in the present study. For this purpose it is recommended to study the dynamics of the ratio function $r$:

\be
r=\frac{\rho_m}{\rho_\phi}=\frac{\Omega_m}{\Omega_\phi},\label{ratio}\ee
in respect to the time variable $\tau\equiv \ln a$ (it is related to the cosmic time through $d\tau=H dt$). The following generic evolution equation holds for $r$:

\be r'=\frac{\rho_m}{\rho_\phi}(\frac{\rho_m'}{\rho_m}-
\frac{\rho_\phi'}{\rho_\phi})=f(r),\label{evolutionr}\ee where the prime denotes derivative in respect to $\tau$, and $f$ is an arbitrary function (at least of class ${\cal C}^1$) of $r$. One is then primarily interested in the equilibrium points of equation (\ref{evolutionr}), i.e., those points $r_{ei}$ at which $f(r_{ei})=0$. After that, one expands $f$ in the neighborhood of each equilibrium point; $r=r_{ei}+\epsilon_i$, so that, up to terms linear in the perturbations $\epsilon_i$: $f(r)=(df/dr)_{r_{ei}}\epsilon_i+{\cal O}(\epsilon_i)\Rightarrow \epsilon_i'=(df/dr)_{r_{ei}}\epsilon_i$. This last equation can be integrated to yield the evolution of the perturbations in time $\tau$:

\be
\epsilon_i=\epsilon_{0i}\exp{[(df/dr)_{r_{ei}}\tau]},\label{perts}\ee
where $\epsilon_{0i}$ are arbitrary integration constants. It is seen from (\ref{perts}) that, only those perturbations for which:

\be (df/dr)_{r_{ei}}<0,\ee decay with time $\tau$, and the corresponding equilibrium point is stable. The necessary condition to evade the CCP is then given by the requirement that the point $\rho_m/\rho_{\phi}=r_{ei}\lesssim 1$ be stable against small linear perturbations of the kind explained above.

If we take into account the conservation equations (\ref{densityconservation}) and (\ref{scalarconservation}), and the definition of the interaction term $Q$ given in equation (\ref{interactionfunction}), then, the function $f$ can be given by the following expression:

\be
f(r)=r\{(\ln\bar\chi)'(r+1)+3(\gamma_{\phi}-\gamma_m)\}.\label{function}\ee
Note that, for a model without interaction ($(\ln\bar\chi)'=0$) and with a constant DE barotropic parameter $\gamma_{\phi}=\gamma_{\phi,0}$ (consider, for simplicity, dust-like background fluid so that $\gamma_m=1$); $f(r)=3(\gamma_{\phi,0}-1)\;r$ and the only (stable) equilibrium point is the dark energy dominated solution $r=0\Rightarrow \Omega_\phi=1$. In consequence the coincidence does arise in this case.

Now we modify the method used before to generate FRW solutions, by considering the necessary condition to avoid the CCP. The idea is to give the functions $f=f(r)$ and $\gamma_\phi=\gamma_\phi(r)$ as the input functions instead of $F$ and the coupling function $\bar\chi$. Note, in this regard that, according to (\ref{evolutionr});

\be d\tau=\frac{da}{a}=\frac{dr}{f(r)},\label{tauvar}\ee then the scale factor (and any other relevant function) can be written as function of the ratio $r$:

\be a(r)=\alpha \exp{\int\frac{dr}{f(r)}},\label{ar}\ee where $\alpha$ is an arbitrary integration constant. The next step is to choose a function $f$ such that, at least, one of the ruts $r_e$ of equation $f(r)=0$; obeys that $(df/dr)_{r_e}<0$. As explained above, this is the necessary requirement for a given solution to be free of the cosmic coincidence.

Note that, after considering (\ref{function}) and (\ref{tauvar}), the equation (\ref{evolutionr}) can be rearranged in the following way:

\be \frac{d\ln\bar\chi}{dr}=\frac{1}{r(r+1)}-
\frac{3(\gamma_\phi-\gamma_m)}{(r+1)f(r)},\ee or, after
integration:

\be \bar\chi(r)=\bar\chi_0(\frac{r}{r+1})\exp{[-3\int\frac{dr(\gamma_\phi-\gamma_m)}{(r+1)f(r)}]},\label{barchi}\ee where, as before, $\bar\chi_0$ is an arbitrary integration constant, and we
have considered that the barotropic index $\gamma_\phi$ can be given as function of the ratio $r$: $\gamma_\phi=\gamma_\phi(r)$. This means that, once $f(r)$ and $\gamma_\phi(r)$ are given as input functions, then, by means of equation (\ref{barchi}) we can find the coupling function $\bar\chi=\bar\chi(r)$ (or, after (\ref{ar}), it can be given as function of the scale factor:
$\bar\chi=\bar\chi(a)$). This means, in turn, that $\rho_m$ (equation (\ref{density})) and $\rho_\phi=\rho_m/r$ are known functions, i.e., the function $G(r))$ (equation (\ref{Gfunction})) is also known. On the other hand, the function $L=L(r)$ (equation
(\ref{Lfunction})) can also be written in the following way:

\be
L(r)=\frac{\epsilon}{2}\gamma_\phi(r)\rho_\phi(r),\label{Lr}\ee
so, once $\gamma_\phi(r)$ is given as input, $L=L(r)$ is also known. Now, since $G$ and $L$ are both known, then we are able to derive exact solutions in quadratures through the integrals
(\ref{deltaT}), and (\ref{deltaFi}), that can be rewritten as follows:

\be \Delta
t=\pm\sqrt{3}\int\frac{dr}{f(r)\sqrt{G(r)}},\label{deltaTr}\ee and

\be \Delta\phi=\pm\sqrt{6}\int\frac{dr}{f(r)}
\sqrt\frac{L(r)}{G(r)},\label{deltaFir}\ee respectively.

Recall that, once the relevant magnitudes are given as functions of the ratio $r$, these can be given as functions of the scale factor also, through either relationship (\ref{tauvar}) or (\ref{ar}).

\subsection{CCP-free Model.}\label{soluA}

As it was stated in section \ref{coincidence} it is possible to construct cosmological models which are free from the coincidence problem by selecting apropiated functions $f=f(r)$ and $\gamma_\phi=\gamma_\phi(r).$ In the case we shall discuss here, we chose:

\be
f(r)=\beta(r_0-r),\;\;\gamma_\phi=\gamma_{\phi,0},\label{example}\ee
where $\beta$ and $r_0$ are positive reals, meanwhile, the constant $\gamma_{\phi,0}$ could be either positive or negative. Recall that the solution comprises both quintessence $(0\leq\gamma_{\phi,0}\leq2/3)$ and phantom $(\gamma_{\phi,0}<0)$ behavior. Notice that, if one substitutes the function $f(r)$ given above, in the evolution equation for the ratio $r$ eq. (\ref{evolutionr}), then the equilibrium point $r_e=r_0$ is stable. Therefore, the coincidence problem is avoided whenever $r_0\lesssim 1$. 

Through (\ref{ar}) the relationship between the scale factor and the ratio $r$ can be readily found:

\be a(r)=\frac{\alpha}{(r-r_0)^{1/\beta}}.\label{eje1}\ee Whereas, the coupling function looks like:

\be \bar\chi(r)=\bar\chi_0
\frac{r}{r+1}\left(\frac{r-r_0}{r+1}\right)^{3\frac{\gamma_{\phi,0}-\gamma_m}{\beta(1+r_{0})}},
\label{exe1coupfunt} \ee

From equation (\ref{Gfunction}), we find the expression for $G$:

\be G=G_0 \left[\alpha(r-r_{0})\right]^{3\gamma_{m}/\beta}
\left(\frac{r-r_0}{r+1}\right)^{3\frac{\gamma_{\phi,0}-\gamma_m}{\beta(1+r_{0})}}
,\label{G1} \ee where $G_0=\bar\chi_0 \rho_{m0},$  besides, from equation (\ref{Lr}) the following expression for $L$ is obtained:

\be L=G_0 \gamma_{\phi,0}
\frac{\left[\alpha(r-r_{0})\right]^{3\gamma_{m}/\beta}}{2 r}
\frac{r}{r+1}\left(\frac{r-r_0}{r+1}\right)^{3\frac{\gamma_{\phi,0}-\gamma_m}{\beta(1+r_{0})}}.\label{L1} \ee

Using Eqs. (\ref{G1}, \ref{L1}, \ref{deltaTr}) and eq. (\ref{deltaFir}), $\Delta t$ and $\Delta \phi$ can be computed to give:

\bea &\Delta t=\pm\frac{2}{A \beta}\sqrt{\frac{3}{G_0}} (r_0+1)^{-B/2}(r-r_0)^{-A/2}\times\nonumber\\
&{}_{2}F_1\left[-\frac{A}{2},\frac{B}{2},1-\frac{A}{2},\frac{r_0-r}{1+r_0}\right], \label{dt1}\eea where $A=3[\gamma_m(r_0+1)+1]/[\beta(1+r_0)]$, $B=-3(\gamma_{\phi,0}-\gamma_m)/[\beta(1+r_0)],$ and

\bea
&&\Delta\phi=\pm \sqrt{12\gamma_{\phi,0}(1+r_0)}\beta \{ \sqrt{1+r_0}\; arc\sinh(\sqrt{r})
\nonumber\\
&&-\sqrt{r_0}\;arc\tanh(\sqrt{\frac{r(1+r_0)}{r_0(1+r)}})\}, \label{df1}
\eea respectively. The expression of the DE dimensionless energy density is given by: 

\be \Omega_\phi=\frac{a^{\beta}}{\alpha+(r_0+1) a^{\beta}} \ee 

\begin{figure}[t!]
\begin{center}
\hspace{0.4cm}\includegraphics[width=7cm,height=4cm]{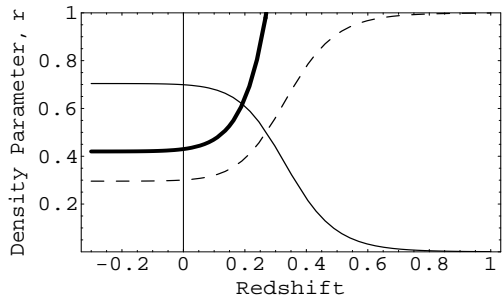}
\end{center}
\caption{A plot of the energy density parameters $\Omega_m$ (dashed curve) and $\Omega_{\phi}$ (thin solid curve), as well as of the ratio function $r$ (thick solid curve) versus redshift, is shown for the model of subsection (\ref{soluA}). For simplicity background dust ($\omega_m=0$) was considered. The values of the free parameters are $\alpha=0.76$, $\beta=17$ and $r_0=0.42$. The ratio function $r$ approaches a constant value $r_0$ at negative redshifts. At present ($z=0$) $r=0.43$, which means that we are already in the long living matter-scaling state. The background energy density dominates the arly stages of the cosmic evolution. At $z\sim 0.28$ both density parameters equate and, since them, the dark energy component dominates the dynamics of the expansion.}
\label{fig1}
\end{figure}

The evolution of the energy density parameters $\Omega_m$, $\Omega_{\phi}$, as well as of the ratio function $r$ versus redshift, is shown in the figure \ref{fig1}. The ratio function $r$ approaches a constant value $r_0$ at negative redshifts. At present ($z=0$) $r=0.43$, which means that we are already in the long living matter-scaling state. The background energy density dominates the early stages of the cosmic evolution. At $z\sim 0.28$ both density parameters equate and, since them, the dark energy component dominates the dynamics of the expansion. At late time, since there is accelerated expansion, the following inequality is always satisfied: $\Omega_\phi>(3\gamma_m-2)/(3\gamma_m-3\gamma_\phi)$. This solution allows for transition from decelerated into accelerated expansion, as required by the well-known cosmological paradigm. 

\section{Crossing of the Phantom Barrier $\omega=-1$}\label{soluB}

The goal of the present section is to study a model which does the crossing of the phantom barrier at late times. To perform the study we use the technique developed in section \ref{coincidence}. 

An incomplete list of previous results on the subject of this section includes the following references: \cite{caldor} (the viability requirements on the equation of state and sound speed are discussed); realizations of the crossing with an extra-dimensional origin, has been discussed in \cite{brane}; proposals in the framework of scalar-tensor theories include \cite{scalar}; \cite{high} (a single field proposal involving high order derivative operators in the lagrangian); \cite{vector} (a model with a single dynamical scalar field coupled to an a priori non-dynamical background covariant vector field). In \cite{chap} an interacting Chaplygin gas is considered. Standar four-dimensional scalar field models are the most popular options of the inventory.

The impossibility of the occurrence of the transition in traditional single field models \cite{theoretical} has motivated much activity in the construction of two field models that do the job. Examples of explicit constructions can be found in \cite{cross}, but perhaps the class of models which have received most attention are quintom cosmologies \cite{quintom,lazkoz}. 

In the present section, as already mentioned, we restrict ourselves to the framework of scalar-tensor theories \cite{scalar}. We choose the following input function and DE equation of state (EOS) parameter $\omega_\phi =\gamma_\phi-1$:

\be
f(r)=\beta (r_0-r),\;\;\omega_\phi=\omega_m+\delta \frac{r+1}{r},\label{example2}\ee respectively. In this equation $r_0$, $\delta$ and $\beta$ are arbitrary constant parameters, and $\omega_m$ is the state parameter of the background fluid. The input function $f(r)$ is the same as in (\ref{example}), so that the CCP may be avoided in the present model as well.

\begin{figure}[t!]
\begin{center}
\hspace{0.4cm}\includegraphics[width=7cm,height=4cm]{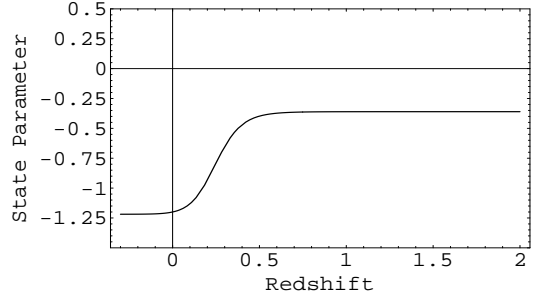}
\end{center}\bigskip
\caption{The DE EOS parameter is plotted as a function of the redshift for the the values $\alpha=0.76$, $\beta=17$, $r_0=0.42$ and $\delta=-0.36$. Note that the DE EOS parameter goes from $\omega_\phi\approx -0.33$ (quintessence) at $z=\infty$, to $\omega_\phi\approx -1.2$ (phantom) at present. The crossing of the $\omega =-1$ barrier is apparent.}
\label{fig2}
\end{figure}

In figure \ref{fig2} the behaviour of the DE EOS parameter vs redshift is shown. The free parameters have been appropriately chosen ($\alpha=0.76$, $\beta=17$, $r_0=0.42$ and $\delta=-0.36$). Notice that $\omega_\phi$ goes from $-0.33$ (quintessence) at $z=\infty$ to $-1.2$ (phantom) at negative redshift. It is apparent that the crossing of the $\omega =-1$ barrier actually happens at $z\sim 0.2$. This is always possible if the dark energy and the dark matter have additional non-gravitational interaction (the coupling scalar field-background fluid is non-minimal). The explanation is simple enough: in standard scalar-field DE models where the coupling is minimal, the crossing can be achieved by adding new degrees of freedom (extra scalar fields). When there is non-minimal coupling instead, there is already an additional degree of freedom that is given by the coupling function in the action (\ref{action}).

Using the same procedure as in the former subsection we find the coupling function:

\be \bar\chi(r)=\bar\chi_0
\left(\frac{r}{r+1}\right)\left(\frac{r-r_0}{r}\right)^{3\delta/(\beta r_0)},
\label{exe2coupfunt} \ee

With this coumpling function we are able to derive exact solutions using the integrals (\ref{deltaTr}) and (\ref{deltaFir}):

\bea
&&\Delta t=\mp
\sqrt{\frac{12 \alpha^{3 \gamma}}{G_0\beta^{2}}}(\frac{r^{1-B/2}(r_0-r)^{A/2}(r-r_0)^{-A/2}}{r_0^{1+A/2}(B-2)})\nonumber\\
&&_{2}F_1 (1-\frac{B}{2}, 1+\frac{A}{2},2-\frac{B}{2},-\frac{r}{r_0}), \label{dt2}
\eea

\bea
&&\Delta\phi=\mp\sqrt{\frac{24((r-1)(1+\omega_m)+(r+1)(1+\omega_m+2\delta))}{\beta^{2}(1+\omega_m+\delta)}}\nonumber\\
&&\{\frac{(1+r_0)(\delta+r(1+\omega_m+\delta))}{\delta+r_0(1+\omega_m+\delta)}\;\Pi (l,-y,k)
\nonumber\\
&&-\frac{(1+\omega_m)}{\delta+r_0(1+\omega_m+\delta)}\; F(y,k)\},\label{df2}
\eea where $y=\arcsin(\sqrt{1/r+1})$, $k=(1+\omega_m)/(1+\omega_m+\delta)$, and $l=1+r_0$. In the above equation $F(y,k)$ is the Legendre elliptic integral of the first kind, while $\Pi (z,y,k)$ is the Legendre elliptic integral of the third kind. 

The dimensionless density parameters have similar behavior as in the former model (figure \ref{fig1}). In consequence the present model is cosnsistent with observational evidence on a matter dominated period in the past and late time DE dominance.

The DM and DE energy densities are given by the following expressions:

\be \rho_m=\rho_0 (\frac{\alpha}{r-r_0})^{-3 \gamma/\beta} \bar\chi_0
\left(\frac{r}{r+1}\right)\left(\frac{r-r_0}{r}\right)^{3\delta/(\beta r_0)}
\label{bde}\ee
\be \rho_\phi=\rho_m/r
\label{de}\ee
At late time (large $a$), the function $r\rho_\phi\propto\rho_m\propto a^{-3(\gamma+\delta/r_o)}$. This means that, whenever $\gamma+\delta/r_o \geq 0$, the energy density is bounded into the future and, consequently, there is not big rip in the future of the cosmic evolution in the model under study.

\section{Concluding remarks}

In this paper we were able to generalize a method for deriving Friedmann-Robertson-Walker (FRW) solutions previously developed in \cite{chimento}, to handle models with non-minimal coupling between the DE and the DM. New classes of exact solutions comprising both phantom and quintessence solutions can be readily derived. 

Interacting models of dark energy are useful to account for the coincidence problem. We took advance of this fact to extended the method of reference \cite{chimento} to derive solutions that are free of the CCP. It has been shown that if an adequated input function $f(r)$ and a barotropic parameter $\gamma_\omega$ are chosen, it is always possible to obtain solutions that avoid the coincindense problem. Besides, due to an extra degree of freedom that is related with the coupling function in the action (\ref{action}), the crossing of the phantom barrier is indeed possible. Although only two specific models were studied in detail, we think it is sufficient to illustrate how the method developed here allows to construct interacting models that avoid the coincidence problem and do the $\omega=-1$ crossing. 

We want to point out that, as shown here, in scalar-tensor models of dark energy where additional non-gravitational interaction between the DE and the DM occurs, it is always possible to find a region in parameter space that makes a given exact solution to be consistent with observations.\footnotemark\footnotetext{A warning has to be made: the evolution of linear perturbations can rule out several models of coupled dark energy \cite{koivisto}, so that in any case a linear pertubations study is mandatory to check the observational relevance of a given model. In the present paper this study is not performed.} It is even possible to avoid the CCP and to find solutions that are free of the big-rip singularity. For this reason, in agreement with previous claims, we conclude that coupled scalar field models are not appropiated to understand the nature of the dark energy at any deeper level \cite{padman}.

\section*{Acknowledgments}

The authors are grateful to B. Gumjudpai, H. S. Zhang, M. Sami, S. Tsujikawa and D. Polarski for calling their attention upon relevant bibliographic references. The MES of Cuba is akcnowledged by partial financial support of the present research.




\end{document}